\documentclass[12pt,preprint]{aastex}
\usepackage[numberedappendix]{emulateapj5}
\usepackage{apjfonts}
\usepackage{amstext}
\usepackage{amsmath}

\newcommand{\nablas}{{\mathbf{\nabla}}\!\!_s}

\newcommand{\mbffone}{{\mathbf {f}}}
\begin{document}
\submitted {Submitted to {\apj} on March 19, 2003}

\journalinfo {Submitted to {\apj} on March 19, 2003}

\shorttitle{RELATIVISTIC MHD AND GRB OUTFLOWS: II.}

\shortauthors{{VLAHAKIS} AND {K\"ONIGL}}

\title{
Relativistic Magnetohydrodynamics with Application to Gamma-Ray Burst Outflows:
II. Semianalytic Super-Alfv\'enic Solutions}

\author{Nektarios Vlahakis and Arieh K\"onigl}
\affil{Department of Astronomy \& Astrophysics and Enrico Fermi
Institute, University of Chicago, 5640 S. Ellis Ave., Chicago, IL 60637}
\email{vlahakis@jets.uchicago.edu, arieh@jets.uchicago.edu}

\begin{abstract}
We present exact radially self-similar solutions of special-relativistic
magnetohydrodynamics representing ``hot'' super-Alfv\'enic outflows
from strongly magnetized, rotating compact objects. We argue
that such outflows can plausibly arise in gamma-ray burst (GRB)
sources and demonstrate that, just as in the case of the
trans-Alfv\'enic flows considered in the companion paper, they
can attain Lorentz factors that correspond to a rough
equipartition between the Poynting and kinetic-energy fluxes and
become cylindrically collimated on scales compatible with GRB
observations. As in the trans-Alfv\'enic case, the initial
acceleration is thermal, but, in contrast to the solutions
presented in Paper I, part of the enthalpy flux is transformed
into Poynting flux during this phase. The subsequent,
magnetically dominated acceleration 
can be significantly less rapid than in trans-Alfv\'enic flows.
\end{abstract}


\section{Introduction}
In the companion paper (\citealt{VK03}, hereafter Paper I) we
presented a general formulation of special-relativistic
magnetohydrodynamics (MHD) and derived exact radially
self-similar solutions for axisymmetric outflows from strongly
magnetized, rotating compact objects.\footnote{In this paper we
continue to use the notation of Paper I; we identify equations, sections, and
figures from that paper by placing the numeral I in front of
their respective numbers.} We pointed out that our
results should be relevant to relativistic-jet sources in a
variety of astrophysical settings (including miniquasars and active
galactic nuclei), but our main focus has been on gamma-ray burst
(GRB) outflows.

In Paper I we concentrated on trans-Alfv\'enic solutions, in
which the poloidal magnetic field component dominates (or at
least is not much smaller than) the azimuthal field
component at the base of the outflow. In the present paper we 
consider outflows in which
$|B_\phi/B_p|\gg 1$ at the origin. The poloidal component of the
Alfv\'en velocity in these flows is very small from the start,
and consequently they are effectively super-Alfv\'enic
throughout.\footnote{The tendency of the Alfv\'en point in
trans-Alfv\'enic outflows to move close to the disk surface when
the value of $|B_p/B_\phi|$ at the base of the flow decreases below $\sim 1$ was illustrated
(in the nonrelativistic context) by \citet{CS94}.} This property
of the flows is demonstrated more formally in \S
\ref{description}. \citet{C95} discussed nonrelativistic, ``cold'' outflows of
this type in the case where $B_p=0$ and $B_\phi\ne 0$. Here we
study super-Alfv\'enic outflows in the relativistic limit, taking
into account thermal-pressure effects and allowing for the
presence of a finite poloidal field component (which is required
for angular-momentum transport by the flow).

As in Paper I and in \citet{VK01}, we consider a debris disk
around a black hole with a massive-star progenitor as a
specific example of a long-duration GRB source. If the outflow emanates from a
region of characteristic cylindrical radius $\varpi_i$ and radial width $(\Delta \varpi)_i$ (where
the subscript $i$ denotes quantities evaluated at the disk surface), and if the
disk injects energy into the flow mostly in the form of a Poynting flux
over a time interval $\Delta t$, then the total injected energy
(from the two surfaces of the disk) is
\begin{equation}\label{E_i}
{\cal E}_i \approx c E_i B_{\phi\,i} \varpi_i (\Delta \varpi)_i
\Delta t\, ,
\end{equation}
where $E=B_p V_\phi/c - B_\phi V_p/c$ is the electric field amplitude. In the trans-Alfv\'enic
case $E \approx B_p V_\phi/c$, and one infers
\begin{eqnarray}\label{B_p}
B_{p\,i} \approx && 3\times 10^{14} \left [  \left (\frac{{\cal E}_i}{10^{52}\, {\rm
ergs}}\right ) \left ( \frac{\varpi_i}{1.6\times 10^6\, {\rm
cm}} \right )^{-2} \left ( \frac{\varpi_i}{2(\Delta \varpi)_i} \right )
\right. \nonumber \\ &&
\left. \times \left ( \frac{V_{\phi\, i}}{10^{10}\, {\rm cm\,
s}^{-1}}\right )^{-1} \left ( \frac{\Delta t}{10\, {\rm s}} \right )^{-1}
\left ( \frac{B_{p\,i}}{|B_{\phi\,i}|}\right ) \right ]^{1/2}\ G\, ,
\end{eqnarray}
where the factor $(B_{p\,i}/|B_{\phi\,i}|)$ on the right-hand side
is typically $\gtrsim 1$ for trans-Alfv\'enic flows. 

A massive star could give rise to a black-hole/debris-disk
system in at least three ways: (1) a prompt
(within $\sim 1\, {\rm s}$ of core collapse) formation in which
no supernova shock is produced (the ``Type I collapsar'' scenario;
e.g., \citealt{MW99}); (2) a delayed (within $30-3000\, {\rm s}$
of core collapse) formation resulting from fallback in a failed supernova
explosion (the ``Type II collapsar'' scenario; e.g., \citealt{F99});
(3) a strongly delayed (by a period of weeks to years) formation
resulting from the collapse (following the loss of rotational
support through a pulsar wind) of a rapidly rotating, massive
neutron star created in a successful supernova explosion (the
``supranova'' scenario; e.g., \citealt{VS98}). In the latter
scenario, the initial magnitude of $B_p$ in the debris
disk would not exceed typical radio-pulsar values ($\sim
10^{12}-10^{13}\, G$), and in the Type I collapsar picture the
magnitude of an inward-advected large-scale poloidal field would likely
be even lower. In the Type II collapsar scenario it is in
principle possible for a strong poloidal field to develop on the
scale of the stalled supernova shock on account of either
convection \citep[e.g.,][]{TM01} or a hydromagnetic instability
\citep[e.g.,][]{A03}, but the ability of these mechanisms to
produce ordered disk poloidal fields of the required magnitude
remains questionable. In any case, most of the long-duration
GRBs observed to date are interpreted in the collapsar picture as
being of the Type I variety \citep{MWH01}.

The most promising mechanism of creating strong fields in debris disks
is through shear amplification of the seed poloidal field by the
strong (Keplerian) differential rotation
\citep[e.g.,][]{MR97}. This process generates azimuthal
fields that can greatly exceed the poloidal field component inside
the disk. \citet{KR98} and \citet*{RTK00}, who 
considered this mechanism in the contexts of both a debris disk
and a newly formed neutron star, suggested that the
wound-up field would rise buoyantly but not necessarily
axisymmetrically and could emerge from the disk (or stellar)
surface as a flux rope with $|B_{p\,i}|$ of order the amplified
$|B_\phi|$. A similar scenario has been proposed for the formation
of active regions in the sun and has been tested by numerical
simulations \citep[e.g.,][]{BCC02,AF03}. \citet{KR98} and
\citet{RTK00} also suggested that repeated episodes of field
windup and escape could be the origin of
the observed intrinsic GRB variability (see \S~I.4.1.1).

\citet{KR98} pointed out that an alternative outcome of the
field winding process might be the escape of disconnected
(through magnetic reconnection) toroidal flux loops. In this
case $|B_{\phi\,i}/B_{p\,i}|$ could be $\gg 1$. This is similar to the ``plasma gun''
scenario proposed by \citet{C95}. \citet{STM90} argued, on the basis of numerical
simulations, that this behavior is likely to occur
explosively and could be manifested by disks in which the
pressure of the amplified field increases above the thermal
pressure (see also \citealt{HTS92} and \citealt{CD94}). It is
interesting to note that the field windup and
reconnection cycle underlying these models was also at the basis of early models of disk
magnetic viscosity \citep[e.g.,][]{EL75,C81,SR84}. More recent
disk-viscosity models have been based on MHD turbulence induced by the magnetorotational
instability \citep[MRI; e.g.,][]{BH98}. The debris
disks invoked in GRB source models are envisioned to be initially threaded by a subthermal
poloidal magnetic field, which could make them
susceptible to this instability. One might then wonder whether
most of the energy liberated in the disk would be viscously
dissipated into heat, with little left to power Poynting flux-dominated
outflows. Although a definitive answer to this question would
entail a fully global numerical investigation of the nonlinear development of
the MRI, existing (radially localized) simulations
\citep[e.g.,][]{MS00} indicate that disks threaded initially by
a weak poloidal field with a nonvanishing mean evolve into
configurations that are magnetically dominated throughout
(and thus no longer MRI-unstable).

In this paper we explore the possibility that at least some
GRB outflows are produced with $|B_{\phi\,i}/B_{p\,i}|\gg 1$. In
this case $E \approx -B_\phi V_p/c$, and equation (\ref{E_i}) implies
\begin{eqnarray}\label{B_phi}
|B_{\phi\, i}| \approx && 3\times 10^{14} \left [  \left (\frac{{\cal
E}_i}{10^{52}\, {\rm ergs}}\right ) \left ( \frac{\varpi_i}{1.6\times 10^6\, {\rm
cm}} \right )^{-2} \left ( \frac{\varpi_i}{2(\Delta \varpi)_i}
\right ) \right. \nonumber \\ &&
\left. \times \left ( \frac{V_{p\, i}}{10^{10}\, {\rm cm\,
s}^{-1}}\right )^{-1} \left ( \frac{\Delta t}{10\, {\rm s}} \right )^{-1} \right ]^{1/2}\ G\, ,
\end{eqnarray}
where $V_{p\,i}$ is estimated to be of the order of the speed of
sound ($~\sim c/\sqrt{3}$) at the base of the
flow. Although the behavior of $B_{\phi\,i}$ in the field-windup
scenario is inherently nonsteady, it can nevertheless be incorporated into the
quasi-steady ``frozen pulse'' formalism presented in Paper I, as
we proved in \S~I.2.1.

A brief overview of the application of the $r$
self-similar model outlined in Paper I to super-Alfv\'enic
flows is given in \S~\ref{description}. We present representative ``hot'' and
``cold'' solutions in \S~\ref{results} and discuss analytic
approximations to our results in \S~\ref{discussion}. Our
conclusions are summarized in \S~\ref{conclusion}.

\section{Outflow Description}\label{description}
As in Paper I, we consider a multiple-shell outflow consisting of
baryonic matter, electrons (that neutralize the protons),
$e^\pm$ pairs in thermodynamic equilibrium with photons,
and a large-scale electromagnetic field. We utilize the
$r$ self-similar, ideal-MHD model described in \S~I.3. Adopting
as typical values $|B_{\phi\,i}| \approx 3\times  10^{14}\, G$ and $B_{p\,i}
\approx 3\times 10^{12}\, G$, we have $x = E/B_p \approx -B_\phi V_p/c
B_p \approx 60 \gg 1$ at the base of the flow, 
so the condition $M^2+x^2>1$ for super-Alfv\'enic flows (see \S~I.3.2.1) is satisfied.
In fact, for typical parameters the ``Alfv\'enic'' Mach number $M$
(eq. [I.15]) is also $\gg 1$ at the base of the flow, which
further sharpens the above inequality. 

From equation (I.9) $V_\phi = \varpi \Omega - V_p (-B_\phi / B_p)$. At the base of a
trans-Alfv\'enic flow $B_p \gtrsim -B_\phi$ and hence $\Omega
\approx V_\phi/\varpi$. However, when $B_p \ll -B_\phi$, the
last term on the right-hand side of the expression for $V_\phi$ is $\gg c$, so the first term
on the same side must also be $\gg c$ in order for their
difference to remain $<c$.\footnote{ 
The implied inequality $\varpi \Omega > c$ is another manifestation
of the fact that the region where $B_p \ll -B_\phi$ is located downstream
from the ``light cylinder,'' and hence is super-Alfv\'enic.}
This implies that $\Omega \approx
-B_\phi V_p/(\varpi B_p)\gg V_\phi/\varpi$ at the base of a super-Alfv\'enic
flow. In contrast to the trans-Alfv\'enic situation, in this
case the fieldline constant $\Omega$ cannot be identified with the matter angular
velocity at the footpoint of the fieldline at the midplane of
the disk. As was already remarked upon by \citet{C95}, the fieldline
angular velocity $\Omega$ is generally not a meaningful material
property. We prefer instead to interpret it in terms of the
electric potential $\Phi$: ${\mathbf{E}} = -({\Omega}/{c}) \nablas A = -({1}/{c}) \nablas \Phi$,
so $\Omega \equiv ({\partial}/{\partial A}) \Phi(A\,,s)$.

In super-Alfv\'enic flows, $E \approx - B_\phi V_p/c \approx
-B_\phi$ everywhere above the disk. This means that the transfield components of
the electric and magnetic forces are comparable, ${\mbffone}_{E \bot}
\approx -{\mbffone}_{B \bot}$. This, in turn, is
conducive to a successful evolution of the outflow, as the
strong magnetic collimation that tends to pinch the flow is
countered by the decollimating electric force, which allows the
jet to reach large distances.\footnote{For $V_{p\,i}\ll c$,
corresponding to $E\ll -B_\phi$, we obtain nonrelativistic
solutions, which resemble the ones found by \citet{C95}.}
\begin{table*}
\begin{center}
\caption{
Parameters of Representative Solutions$^\dagger$
\label{table1}}
{\footnotesize
\begin{tabular}{cccccccccc} 
\tableline\tableline
solution & $F$ & $\theta_i(\degr)$ & $\Theta_i$ &
$\rho_{0\,i}$(g cm$^{-3}$) & $B_{p\,i}$(G) & $B_{\phi\,i}$(G) &
$\vartheta_i(\degr)$ & $V_{p\,i}/c$ & $V_{\phi\,i}/c$
\\
\tableline
$a$ & $1.01$ & $85$ & $0.7685$ & $726$ & $1.2 \times 10^{12}$ &
$-1.2 \times 10^{14}$ & $60$ & $0.7627$ & $0.5$ \\
$b$ & $1.01$ & $85$ & $0.011$ & $437$ & $1.2 \times 10^{12}$ &
$-1.2 \times 10^{14}$ & $60$ & $0.7737$ & $0.5$ \\
\tableline
\multicolumn{10}{l}{
$^\dagger$ In both cases $\Gamma = 4/3\,, z_c=0$,
$\varpi_{i \,, {\rm in}}=1.5 \times 10^6\, {\rm cm}$,
and $r_{{\rm out}}/r_{{\rm in}}=3$.} \\
\end{tabular}
}
\end{center}
\end{table*}

\begin{figure*}
\centering
%  {\includegraphics[width=.8\textwidth]{f1.eps}}
  {\includegraphics[width=.8\textwidth,height=.63\textheight]{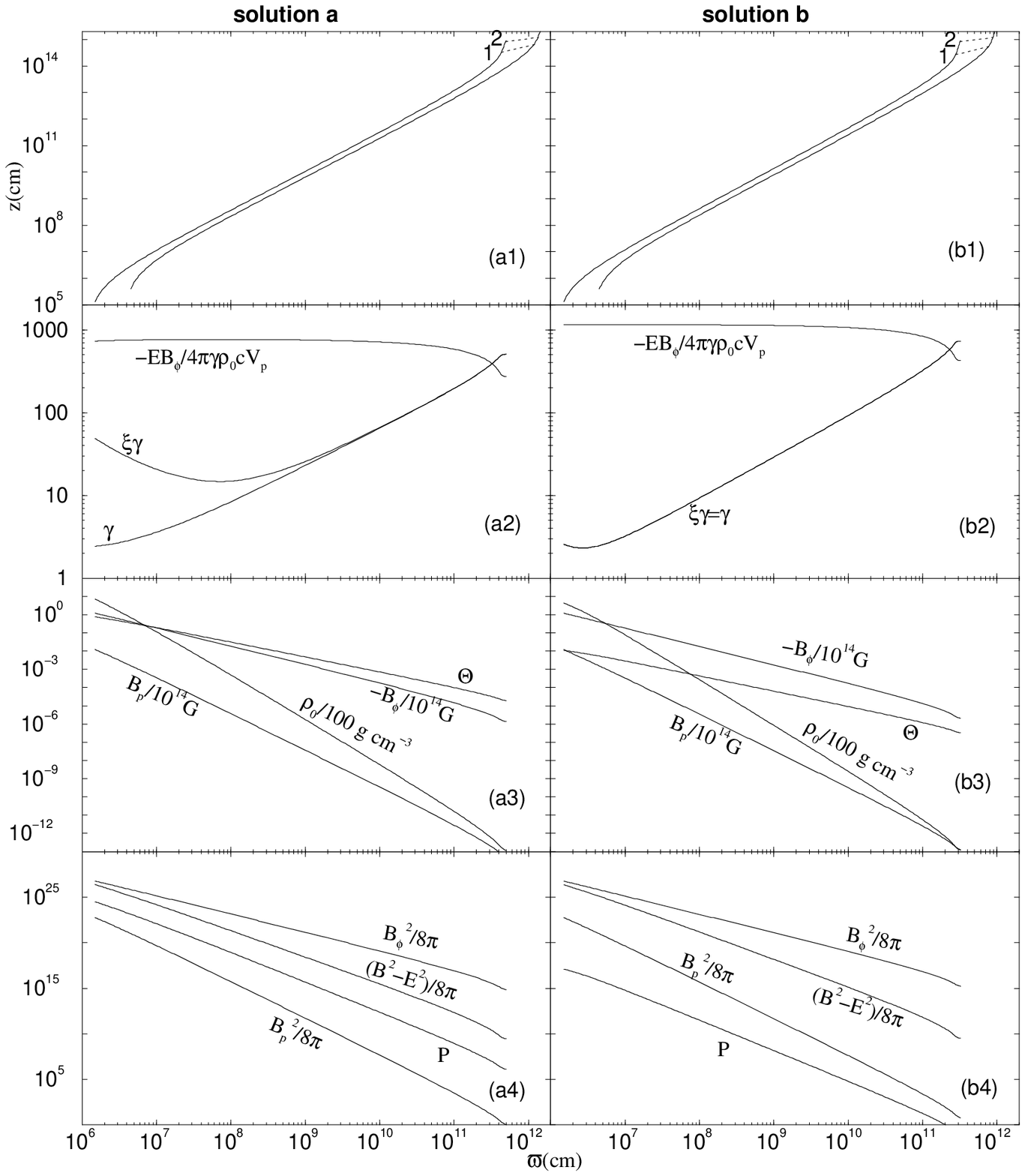}}

\hspace{5mm}
%  {\includegraphics[width=.78\textwidth]{f2.eps}}
  {\includegraphics[width=.78\textwidth,height=.18\textheight]{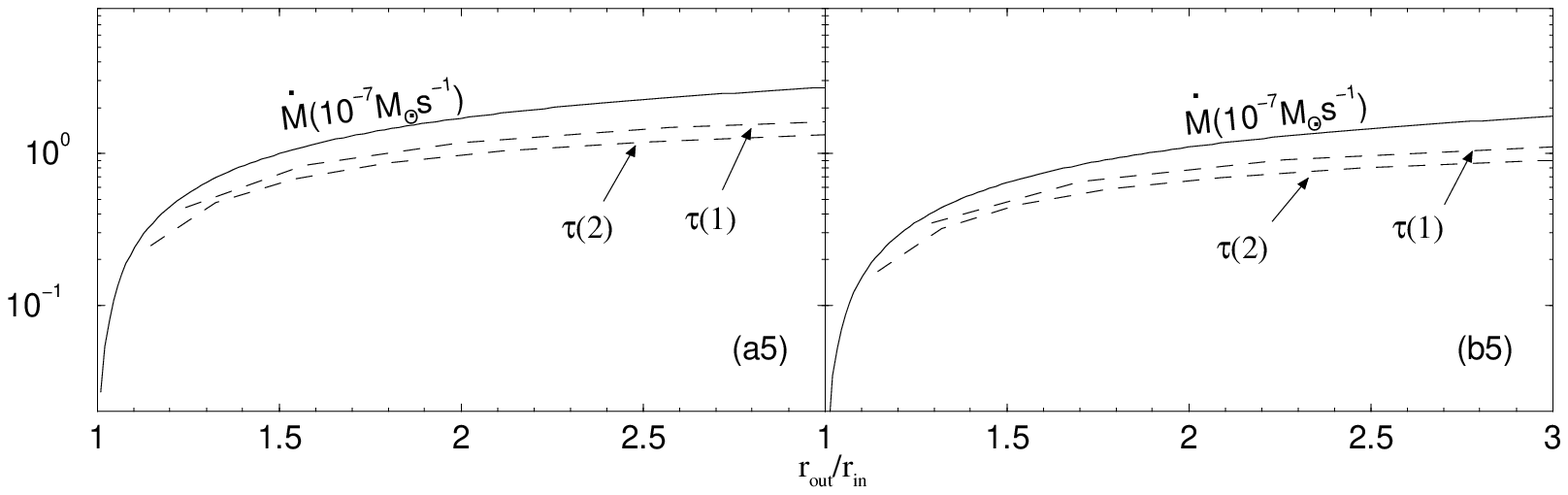}}
\caption{
Main properties of the two representative solutions discussed 
in \S \ref{results}. See text for details.
\label{figall}}
\end{figure*}

The expressions for the various physical quantities that appear
in the $r$ self-similar model can be simplified by taking
account of the inequalities $M^2 \gg 1-x^2$ and  $x\gg x_A$,
which apply in the super-Alfv\'enic regime. We present the
results in Appendix \ref{appA}, showing explicitly the
dependences on $r$, $\theta$, and $s$. For example, the
expression for $B_\phi$ has the form
\begin{eqnarray}
B_\phi = \underbrace{C}_{const} \times
\underbrace{r^{F-2}}_{r} \times \underbrace{
\frac{(\sin \theta)^{F-2} }{{\cal X}^{F-3} ({\cal X}^2 +{\cal M}^2)} 
}_{\theta} \times 
\underbrace{\left[g(ct-\ell)\right]^{1/2}}_{s=ct-\ell} \,.
\nonumber
\end{eqnarray}
Because of the $s$ dependence, the outflow looks like an
outward-propagating wave, which is the generalization of the
electromagnetic outflows (ignoring inertia) described
by \citet{LB02}.\footnote{Their force-free formulation enabled
them to also include the oppositely-directed [$\propto g(ct+\ell)$]
wave, which, however, is $\gamma^2$ times weaker than the
outgoing one.}
The unknown functions of $\theta$ can be obtained after integrating the
ordinary differential equations (I.B2d) and (I.B2e).
For a specific solution we need to give the value of the model parameter $F$
and seven boundary conditions at the base of the flow. The
% I changed ` on the ejection surface of the disk 
latter could be, for example, the values of $\Theta_i\,, \rho_{0\,i}\,, B_{p\,i}\,, B_{\phi\,i}\,, 
\vartheta_i\,, V_{p\,i}\,, V_{\phi\,i}$ at the point ($r_i\,,\theta_i$)
where the innermost fieldline is anchored. (The streamline constants $\mu\,, q\,, \sigma_M \,, x_A$, and
$B_0 \varpi_0^{2-F}$ can then be inferred from eqs. [I.B7a]--[I.B7c].)
The requirement that the outflow cross the modified 
fast-magnetosonic
singular surface imposes a constraint on these boundary conditions: this
is found numerically as in Paper I.

\section{Results}\label{results}

We present the results of the numerical integration for two
representative solutions (labeled $a$ and $b$),  
for which the boundary conditions are given in Table \ref{table1}.
The most important physical quantities are plotted in Figure \ref{figall},
in which each column corresponds to a given solution.
The properties of these solutions are described in
detail in the following subsections. 

\subsection{Solution $a$: A Hot, Fast-Rotator Outflow}
This solution is intended to represent a realistic GRB outflow:
the total energy initially resides predominantly in the electromagnetic field,
but the thermal part is nonnegligible, corresponding to temperatures
$\Theta_i \lesssim 1$.

Figure \ref{figall}a1 shows the meridional projections of the innermost
(anchored at $\varpi_{i \,, {\rm in}}=1.5 \times 10^6\, {\rm cm}$)
and outermost (anchored at $\varpi_{i \,, {\rm out}}=
4.5 \times 10^6\, {\rm cm}$) fieldlines.
Over much of the flow the line shape is fitted by
$(z/ 10^5 \, {\rm cm}) \approx 4 (\varpi /10^6 \, {\rm cm})^{1.48}$,
and it becomes cylindrical asymptotically.
The dashed lines labeled 1 and 2 represent the optical paths
for photons that originate at two points on the innermost streamline.
The corresponding optical depths $\tau(1)$ and $\tau(2)$ are shown in
Figure \ref{figall}a5. It is seen that the flow becomes
optically thin (with $\tau$ decreasing to $\sim 1$) just before reaching
the asymptotic cylindrical regime, corresponding to maximum baryon loading \citep[see][]{VK01}.

Figure \ref{figall}a2 shows the Lorentz factor $\gamma$ and the two parts
of the energy flux in units of $\gamma \rho_0 c^2 V_p$
(i.e., the mass-flux $\times c^2$):
the Poynting contribution $-E B_\phi/ 4 \pi \gamma \rho_0 c V_p$
and the matter contribution $\xi \gamma$. The ratio of these two
parts is the magnetization function
$\sigma= -E B_\phi/ 4 \pi \xi \gamma^2 \rho_0 c V_p$,
whereas their sum is a constant: $\mu = \xi \gamma - E B_\phi/ 4 \pi \gamma \rho_0 c V_p$
(see eq. [I.13d]).\footnote{This constant may, however be different 
for different shells, i.e., $\mu=\mu(s)$.}
All quantities are given as functions of the cylindrical distance $\varpi$
along the innermost streamline. It is seen that $\gamma$
increases continuously; for a significant portion of the flow
it can be fitted by
$\gamma \approx (\varpi /10^6 \, {\rm
cm} )^{0.46}$, and its asymptotic value is $\gamma_\infty \approx 500$.
The magnetization function decreases from $\sigma_i\approx 15$ near the disk
to $\sigma_\infty \approx 0.5$ asymptotically.

Figure \ref{figall}a3 shows the poloidal and azimuthal components of the magnetic field
($B_p \propto \varpi^{-2}$, $B_\phi \propto \varpi^{-1}$),
the temperature ($\Theta \propto \varpi^{-0.813}$), and the comoving baryon density
($\rho_0 \propto \varpi^{-2.4}$).

Figure \ref{figall}a4 shows the thermal pressure (due to
radiation and pairs) $P$, and various
electromagnetic pressures. The displayed solution corresponds to the case
where $B_p^2/8 \pi \ll P \ll B_\phi^2/8 \pi$ throughout the flow.

Figure \ref{figall}a5 shows the radial profile of the mass-loss rate.
For $\varpi_{i \,, {\rm out}}/\varpi_{i \,, {\rm in}}=r_{\rm out}/ r_{\rm in}=3$,
$\dot{M} \approx 2.7 \times 10^{-7}\, M_{\sun}\, {\rm s}^{-1}$,
corresponding to a total ejected baryonic mass of $M_b \approx 2.7 \times 10^{-6}\, M_{\sun}$
for a typical burst duration of $\Delta t \approx 10\, {\rm s}$,
and to a total injected energy ${\cal E}_i = \mu M_b c^2 \approx
3.5 \times 10^{51}\, {\rm ergs}$.
A fraction $(1+\sigma_\infty)^{-1} \approx 66.7 \%$ of ${\cal E}_i$
is converted asymptotically into baryon kinetic energy
$E_k=\gamma_\infty M_b c^2 \approx 2.3 \times 10^{51}\, {\rm
ergs}$ of the two oppositely directed jets (see
eq. [\ref{efficiency}] below).
\subsubsection{Collimation}

\begin{center}
\plotone{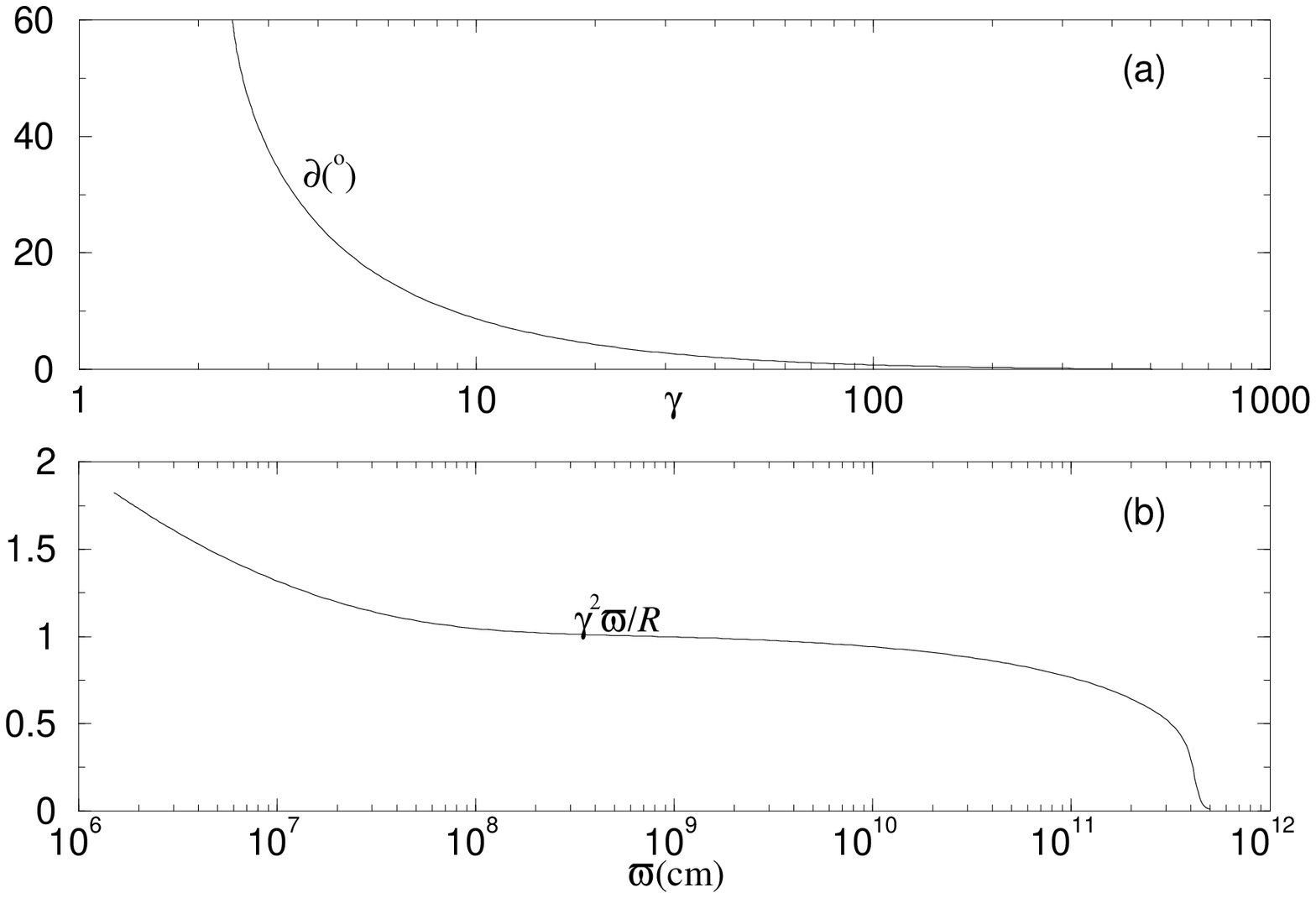}
\figcaption[]
{$(a)$ Opening half-angle of the jet as a function of the Lorentz factor.
$(b)$ The quantity $\gamma^2 \varpi / {\cal R}$ (where ${\cal R}$ is the 
poloidal curvature radius) as a function of $\varpi$. Both figures correspond to solution $a$.
\label{figcollimation}}
\end{center}

It is generally much more difficult to achieve collimation in
the case of relativistic flows than in the nonrelativistic case, both because the
electric force is as large
as the transverse magnetic force and almost cancels it
\citep*[e.g.,][]{CLB98,V03} and because the effective matter inertia is larger \citep[e.g.,][]{B01}.
However, for a flow that is ejected with a mildly relativistic velocity, the
bulk of the collimation can take place before it becomes extremely relativistic.
Figure \ref{figcollimation}a shows that most of the collimation
in the displayed solution happens near the disk: the flow starts
with $\vartheta_i = 60\degr$ and $\gamma_i = 2.4$ at $\varpi_{i
\,, {\rm in}}=1.5
\times 10^6\, {\rm cm}$, but by $\varpi\approx 10^8\, {\rm cm}$,
where $\gamma \approx 10$, it is already collimated to
$\vartheta \approx 10\degr$. 
The collimation is completed at larger values of $\gamma$, and
the streamlines eventually become cylindrical.
Figure \ref{figcollimation}b shows that for $\varpi \gtrsim
10^8\, {\rm cm}$ (corresponding to $\gamma \gtrsim 10$) and up
to $\varpi \approx 10^{11}\, {\rm cm}$ (where the cylindrical regime commences),
the quantity $\gamma^2 \varpi / {\cal R}$ is approximately constant (\citealt{CLB98}; see 
also \S~\ref{discussion}). Despite the large value of the curvature
radius, it is still possible for the opening half-angle to vanish asymptotically.

The collimation is the result of the interplay between the 
magnetic and electric forces in the transfield direction.
Figure \ref{figforces}b shows all the force components in that direction.
It is seen that both ${\mbffone}_{E \bot}$ and ${\mbffone}_{B \bot}$ change sign
(at $\varpi \approx 5\times 10^6\, {\rm  cm}$ and $7\times
10^6\, {\rm cm}$, respectively).
In a small region very close the origin ${\mbffone}_{E \bot}>0$
(corresponding to a positive charge density, $J^0>0$) and ${\mbffone}_{B \bot}<0$
(corresponding to the return-current regime $J_{\parallel}>0$).
In this region the electric force collimates and the magnetic
force acts to decollimate the flow. At larger distances ${\mbffone}_{E \bot}<0$
(the charge density becomes negative, $J^0<0$) and ${\mbffone}_{B \bot}>0$ 
(corresponding to the current-carrying regime $J_{\parallel}<0$).
Beyond $\varpi \approx 10^8$cm these two forces become almost equal
(their difference is the much smaller force
$-{\mbffone}_{I \bot}$), and the collimation continues slowly.

\begin{center}
\plotone{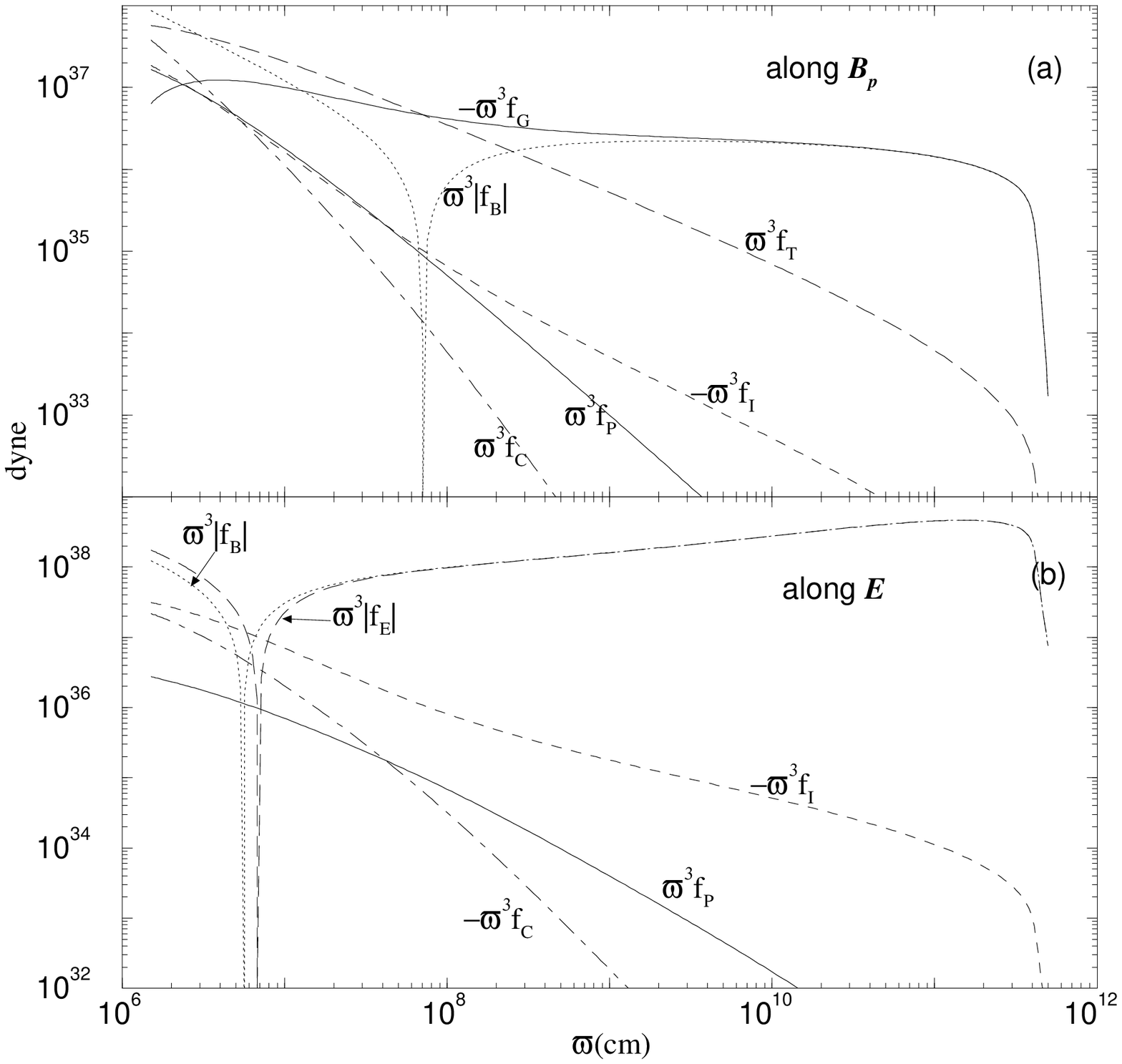}
\figcaption[]
{Force densities (multiplied by $\varpi^3$)
along the poloidal flow ({\emph {top}} curve) 
and in the transfield direction ({\emph {bottom}} curve),
as functions of $\varpi$ along the innermost streamline of solution $a$.
The various force components are defined in eq. (I.18).
\label{figforces}}
\end{center}

\subsubsection{Acceleration}

The poloidal force components along the flow are depicted in Figure
\ref{figforces}a. Referring also to Figure \ref{figall}a2, we
identify three different regimes (as in the corresponding
trans-Alfv\'enic solution described in \S~I.4.1):

1. \underline{Thermal acceleration region}\\
For $\varpi \lesssim 3 \times 10^8\, {\rm cm}$, the thermal
force exceeds the magnetic force, ${\mbffone}_{T \parallel}>{\mbffone}_{B \parallel}$.
The acceleration is thermal: the increase in $\gamma$ is caused
by the decrease in the specific enthalpy $\xi$.\footnote{
There is a very small region near the origin where $-{\mbffone}_{G} $ is small,
meaning that the Lorentz factor remains constant (see also Fig. \ref{figall}a2).
However, even though $V^2$ remains constant, the poloidal
velocity increases (while $V_\phi$ decreases).}
\\
A new feature of this solution (not encountered in the corresponding
trans-Alfv\'enic solution presented in Paper I) is that
%Contrary to the transAlfv\'enic solutions of VK03a, here, 
part of the enthalpy flux in this regime is transformed into Poynting flux:
for $\varpi < 7 \times 10^7\, {\rm cm}$ the ${\mbffone}_{B
\parallel}$ component is negative (corresponding to $J_\bot < 0 $),
resulting in an increase in the Poynting flux and a decrease in the
enthalpy flux (see Fig. \ref{figall}a2).
(In the corresponding regime of solution $a$ of Paper I, the
enthalpy flux  $\propto \xi \gamma$ remains constant.)
The magnetization function increases from $\sigma \approx 15$ near the disk
to $\sigma \approx 50$ at $\varpi \approx 7 \times 10^7\, {\rm cm}$, before
starting to decrease in the regime where ${\mbffone}_{B \parallel}>0$.
\\
Combining the conservation relations for the total specific angular momentum
and the total specific energy (eqs. [I.13c] and [I.13d]), we get
\begin{equation}\label{x_A}
\frac{\Omega}{c^2} \varpi V_\phi = 1 + (x_A^2-1) \frac{\mu}{\xi \gamma}\ .
\end{equation}
Note that for the adopted fiducial values the quantity
\begin{eqnarray}
x_A^2-1 =\frac{x M^2}{M^2 + x^2 - x V_\phi/c} \left(\frac{V_\phi}{c}-\frac{B_p}{E}\right)
\nonumber
\end{eqnarray}
is positive (in contrast to the trans-Alfv\'enic case, where $x_A^2<1$).
Thus, equation (\ref{x_A}) implies that $\xi \gamma$ decreases
with increasing $ \varpi V_\phi$. This explains the behavior of $\xi \gamma$
near the base of the flow and indicates when this effect would
be most pronounced: it may be expected that $\varpi V_\phi$ would increase faster
for a larger initial opening half-angle $\vartheta_i$, so that
the region where ${\mbffone}_{B \parallel}<0$ would be more extended in this case.

2. \underline{Magnetic acceleration region}\\
{}From the end of the thermal acceleration zone, where $\xi \approx 1$, up to
$\varpi \approx 4 \times 10^{11}$ cm, it is seen from Figure \ref{figall}a2
that the Lorentz factor continues to increase.
The acceleration in this regime is due to magnetic effects:
Poynting flux is transformed into kinetic energy flux.
Figure \ref{figforces}a shows that the force $-{\mbffone}_{G \parallel}$ 
(which measures the increase in $\gamma$) is
equal to the magnetic force ${\mbffone}_{B \parallel}$
(which derives from the decrease in $\mid \varpi B_{\phi}\mid $).
\\
Even though $\varpi /{\cal R} \lesssim 1/\gamma^2$ (see Fig. \ref{figcollimation}b),
in agreement with the analysis of \citet{CLB98}, their argument 
that no significant acceleration can take place at large
distances from the light cylinder is evidently circumvented in
this case: the
$\sigma$ function in the magnetic acceleration region
decreases from $\sigma\approx 50$ to the asymptotic value $\sigma_\infty \approx 0.5$.
This behavior is explained in \citet{V03}, who analyzes the efficiency of
magnetic acceleration in relation to the fieldline shape and the role of the
centrifugal force in the transfield force-balance equation.
As the flow becomes progressively more matter-dominated,
an important transition
is reached (at $\sigma=\sigma_{\rm c}$) when the centrifugal force 
becomes equal to the electromagnetic
force component in the transfield direction. The acceleration
could in principle continue beyond that point only if a transition
from a positive to a negative poloidal curvature were possible. 
This cannot happen in $r$ self-similar models, which therefore
have $\sigma_\infty \le \sigma_{\rm c}$.
Large asymptotic cylindrical radii correspond to a comparatively 
smaller centrifugal force and therefore, (on account of
${\mbffone}_{B \bot}+{\mbffone}_{E \bot}=-{\mbffone}_{C \bot}$)
to a less magnetized asymptotic flow.
This explains why the super-Alfv\'enic solutions presented in this paper
attain smaller values of $\sigma_\infty$ 
than the corresponding solutions of Paper I,
which generally collimate more efficiently than super-Alfv\'enic flows.

3. \underline{Asymptotic cylindrical region}\\
At the end of the magnetic acceleration region the
flow becomes cylindrical: Figure \ref{figall}a1 shows that the
fieldlines converge to a constant value of $\varpi$ and that
subsequently all the flow quantities remain constant.

\subsubsection{Time-Dependent Effects}

The disk activity that determines the time variability of a GRB
remains an open question, so one is not yet in a position to accurately
specify the time profile of the boundary conditions at the base
of the flow. We nevertheless present in this
subsection an illustrative example of how one may recover the
time dependence imprinted on the flow by the $s$ dependence of
the boundary conditions at the origin. In particular, we show how
to obtain the function $\gamma(\ell\,, s)$ [or, using $s=ct-\ell$,
the function $\gamma(\ell \,, t)$; see also \S~I.4.1.1].

Equations (\ref{xigamma_s}) and (\ref{xi_s}) give
\begin{equation}\label{pulse1}
\gamma(\ell\,, s)=\gamma(\ell\,, s_0)
\frac{\mu(s)}{\mu(s_0)} 
\frac{\xi(\ell\,, s_0)}{\xi(\ell\,, s)}
\end{equation}
and
\begin{equation}\label{pulse2}
\frac{\xi(\ell\,, s)}{\left[\xi(\ell\,, s)-1\right]^3}=
\frac{g(s)}{g(s_0)}
\frac{q(s_0)}{q(s)}
\frac{\xi(\ell\,, s_0)}{\left[\xi(\ell\,, s_0)-1\right]^3} \ .
\end{equation}
For given $g(s)$, $g(s)/\mu(s)$, $q(s)$ and a known solution 
$\{\xi(\ell\,, s_0) \,, \gamma(\ell\,, s_0)\}$
for the reference shell $s_0$, we can solve this system of
equations to obtain $\{\xi(\ell\,, s) \,, \gamma(\ell\,, s)\}$ for all the other shells
of the outflow.

As a concrete example, we choose the functions $g(s)$ and $g(s)/\mu(s)$ as
shown in the inset of Figure \ref{figpulse}.
In the context of the internal-shock model of GRBs, these
functions can be interpreted as the envelope profiles of a
series of discrete pulses.
Since the Poynting flux is $\propto g(s)$, we choose
$g_{\rm max}/ g_{\rm min} = 10^4$, corresponding to an azimuthal magnetic
field amplification factor $B_{\phi \ \rm max}/ B_{\phi \ \rm min} =10^2$.
The profile of $g(s)/\mu(s) \propto \gamma \rho_0$ is similar to
the one adopted in Paper I. We also set $q(s) \propto g(s)
\mu^2(s)$, for which choice the solution of equations
(\ref{pulse1}) and (\ref{pulse2}) near the origin (where $\xi
\gg 1$) is $\gamma(\ell\approx 0\,, s) \approx \gamma(\ell
\approx 0\,, s_0)$.\footnote{For these forms of the $s$ dependence, 
we verified that the neglected terms in the momentum
equation (I.12e) are very small compared with the other terms of that
equation.} 
The resulting Lorentz factor profile $\gamma(\ell \,, t)$ is shown in Figure \ref{figpulse}.
\begin{center}
\plotone{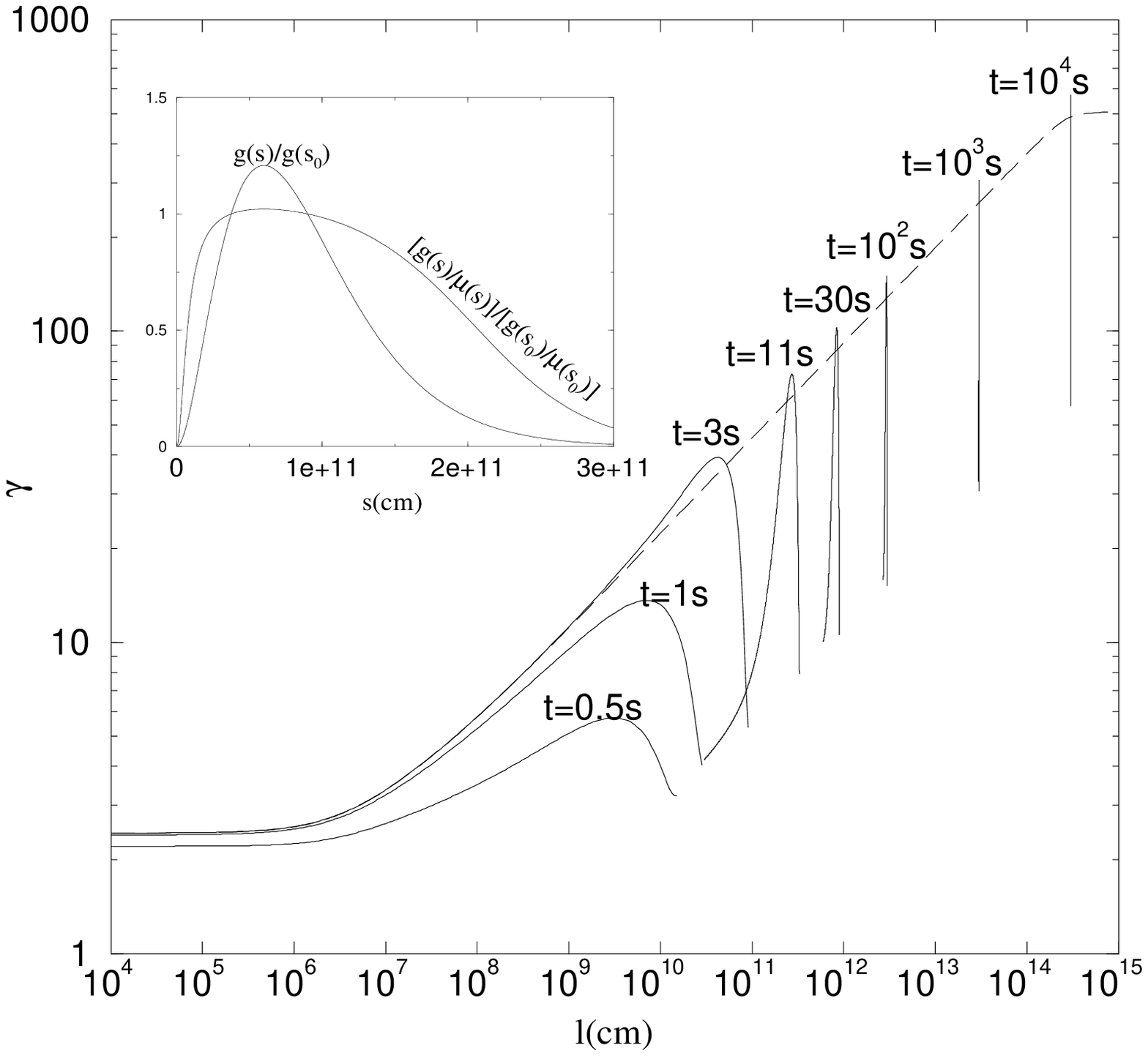}
\figcaption[]
{Lorentz factor plotted as a function of the arclength along the
innermost fieldline of solution $a$ for several values of time
since the start of the burst.
The {\it dashed}\/ line represents the ``time independent''
solution for the reference shell $s_0=0.9 \times 10^{11}$ cm.
The assumed forms of the functions $g(s)$ and $g(s)/\mu(s)$
across the ejected ``pancake'' are shown in the inset.
\label{figpulse}}
\end{center}
For $t=0.5\, {\rm s}$, only the shells with $s \in \left[0\,,\, ct=1.5 \times 10^{10}\mbox{cm} \right]$
have been ejected.
The ejecta occupy the region from $\ell =0$ to $\ell = ct=1.5 \times 10^{10}\, {\rm cm}$, and
the first shells have already been accelerated.

At time $t=11\, {\rm s}$, the whole ``pancake''
$s \in \left[0\,,\, c \Delta t=3 \times 10^{11}\mbox{cm} \right]$
can be seen (assuming a typical duration of $\Delta t = 10\, {\rm s}$).
It occupies the region $\ell \in \left[ct- c \Delta t = 3 \times 10^{10}\, \mbox{cm}
\,,\, c t =3.3 \times 10^{11}\, \mbox{cm} \right]$.

The acceleration continues and at $t = 10^4\, {\rm s}$ the asymptotic stage is reached.
Although the shells were ejected with the same $\gamma_i$, they
now span a wide range of Lorentz factors [corresponding to the linear dependence
$\gamma_\infty (s) = \left(\sigma_\infty+1\right)^{-1} \
\mu(s)$], which would have the effect of enhancing the emission
efficiency of the resulting internal shocks  \citep[e.g.,][]{KPS97}.

\subsection{Solution $b$: The Cold Case ($\xi_i \approx 1$)}

The cold solution presented in the second column of Figure \ref{figall}
is more magnetized than solution $a$: the initial value of its
magnetization function is $\sigma_i \approx 450$.
As a result, it collimates faster and the asymptotic jet radius $\varpi_\infty$ is 
smaller than the corresponding radius in the fiducial solution.

The asymptotic baryon kinetic energy is the same as in solution $a$,
and the baryon loading is similarly the maximum allowed under the
condition that the flow be optically thin in the asymptotic regime ($\tau_\infty \approx 
\sigma_T \varpi_\infty \rho_{0 \ \infty}/m_p \approx 1$).
Since $E_k \propto \rho_{0 \ \infty} \gamma_\infty^2 \varpi_\infty^2 \Delta t
\propto \tau_\infty \gamma_\infty^2 \varpi_\infty$, the smaller asymptotic jet radius
implies a larger value of $\gamma_\infty$ (and hence a smaller
baryonic mass). The actual numbers for solution $b$ are
$\dot{M} \approx 1.75 \times 10^{-7}\ M_{\sun}\ {\rm s}^{-1}$,
$\gamma_\infty \approx 730$ (corresponding to $\sigma_\infty \approx 0.58$),
and an acceleration efficiency $E_k / {\cal E}_i = (1+\sigma_\infty)^{-1}
\approx 63.3 \%$.
In this case $P\ll B_p^2/8 \pi \ll B_\phi^2/8 \pi$ throughout the flow.

Figure \ref{figall}b2 shows a slight initial decrease in $\xi
\gamma \approx \gamma$. This behavior is similar to the one
exhibited by solution $a$ and again corresponds to a negative
value of ${\mbffone}_{B \parallel}$. In this case, however,
the Poynting flux increases not at the expense of the thermal
energy but rather because a fraction of the kinetic energy of
rotational motion is transformed into electromagnetic energy
(with the remainder of
the rotational energy accounting for the increase in the kinetic
energy of poloidal motion in this region).

\section{Discussion}\label{discussion}
The outflow solutions depicted in Figure \ref{figall} exhibit a
power-law behavior over most of their extent. We now show that
these power-law scalings can be obtained analytically.

The Bernoulli equation implies $V_p \approx c$, or, using equation (\ref{V_s}),
\begin{equation}\label{scal1}
\frac{F \sigma_M}{\mu} \frac{\sin \theta}{\sin(\theta-\vartheta)} 
\frac{{\cal X}^2 +{\cal M}^2}{{\cal X}^2} \approx 1\,.
\end{equation}
Well above the disk surface (where $\theta\,, \vartheta \ll \pi
/2$ and $\xi \approx 1$), equation (\ref{scal1}) yields
\begin{equation}\label{scal2}
\left(\frac{d \ln \varpi }{d \ln z} \right)_A = \frac{\vartheta}{\theta } =
1- \frac{F \sigma_M}{\mu} \left(1+\frac{1}{\sigma}\right) \ ,
\end{equation} 
where we approximated the magnetization function $\sigma = -(c E
B_\phi/ 4 \pi)/ (\xi \gamma^2 \rho_0 c^2 V_p)$ by ${\cal X}^2 / {\cal M}^2 $.
In the Poynting flux-dominated regime $\sigma \gg 1$,
so the fieldline shape is
\begin{equation}\label{shape}
z \propto \varpi^{\frac{\mu}{\mu - F \sigma_M}}\, .
\end{equation}
The opening half-angle $\vartheta \propto \varpi^{-F \sigma_M / (\mu - F \sigma_M)}$,
and the curvature radius satisfies
\begin{equation}\label{shape_1}
\frac{\varpi}{ {\cal R} } =
-\varpi \frac{\partial^2 \varpi(z\,, A)}{\partial z^2} \left(\frac{B_z}{B_p}\right)^3
\approx \frac{F \sigma_M}{\mu}\left(1-\frac{F \sigma_M}{\mu}\right) \frac{\varpi^2}{z^2}\ .
\end{equation}

The transfield force-balance equation can be written as
\begin{equation}\label{transfieldapprox}
\frac{\varpi \nablas A} { \mid \nablas A \mid} \cdot \nablas
\ln \mid \frac{\varpi B_{\phi} }{\gamma} \mid 
\approx \gamma^2 
\frac{\varpi}{ {\cal R} }
\frac{B_p^2}{B_\phi^2}
(M^2+x^2)
\end{equation}
\citep[e.g.,][]{CLB91,O02,V03}. 
Using $x^2/M^2 \approx \sigma \gg 1$, 
$-B_\phi \approx E = x B_p$, and assuming that 
$-\varpi B_{\phi} /\gamma = A^{m} {\cal F}(\ell)$
(a form verified by numerical integration at least in the
current-carrying regime),
we get
\begin{equation}\label{gamma_1}
\gamma^2 \frac{\varpi}{ {\cal R} } =m \frac{\varpi \mid \nablas A \mid }{A} \,.
\end{equation}
From the definition of the fieldline constant $\sigma_M$ ($=A \Omega^2 / \Psi_A c^3$) and
the fact that the flow is Poynting flux-dominated
($\mu \approx - \varpi \Omega B_\phi / \Psi_A c^2$) we obtain
\begin{equation}\label{gamma_2}
\frac{\varpi \mid \nablas A \mid } {A} =
\frac{\mu}{\sigma_M}\ ,
\end{equation}
so equation (\ref{gamma_1}) yields
\begin{equation}\label{gamma_3}
\gamma^2 \frac{\varpi}{ {\cal R} } =m \frac{\mu}{\sigma_M} \,.
\end{equation}
Figure \ref{figcollimation}b verifies that the right-hand side
is a constant ($\approx 1$).
Equation (\ref{shape_1}) then implies
\begin{equation}
\gamma \approx
\frac{\mu}{F \sigma_M} \left(\frac{mF}{1-F \sigma_M/\mu}\right)^{1/2}
\frac{z}{\varpi} 
\ ,
\end{equation}
i.e., 
\begin{equation}\label{gamma}  
\gamma
\propto \varpi^\beta\,, \quad
\beta=\frac{F \sigma_M}{\mu - F \sigma_M}\ ,
\quad \mbox{and} \quad \gamma \propto z^{(F \sigma_M / \mu)} \ .
\end{equation}
The scaling $\gamma \propto z/ \varpi$ can also be derived from kinematic 
considerations:
$d \varpi \sim V_\varpi dt \sim (c^2 - V_z^2)^{1/2} dt$,
and for $V_\varpi \ll V_z$ we get $ \gamma \sim  dz/ d \varpi \propto z/ \varpi$ 
for a fieldline shape of the form of equation (\ref{shape}).

The solutions derived in \S \ref{results} have $\beta \approx
0.46$ and $0.49$
in the ``hot'' and ``cold'' cases, respectively.
The other scalings are easier to derive:
using equation (\ref{gamma_2}) we get
\begin{equation}
B_p \approx \frac{\mu A } {\sigma_M} \varpi^{-2}\,, \quad
B_\phi \approx -x B_p \approx -\frac{\mu A \Omega } {c \sigma_M} \varpi^{-1}\,.
\end{equation}
Note that it is, in fact, the deviation from the last two scaling relations that 
gives rise to the acceleration:
\begin{equation}
\frac{\mu A \Omega } {c \sigma_M} +\varpi B_\phi
= \frac{A \Omega } {c \sigma_M} \gamma
\propto \varpi^\beta
\end{equation}
(where we used the definition [I.13d] of $\mu$ in the limit $\xi\simeq 1$).
Finally, from the constancy of the mass-to-magnetic flux ratio
and from the polytropic relation we infer
\begin{equation}
\gamma \rho_0 \approx
\frac{\Psi_A \mu A}{4 \pi c \sigma_M} \varpi^{-2}\,, \quad
\Theta \propto \rho_0^{1/3} \propto \varpi^{-(\beta+2)/3}
\, .
\end{equation}
The fiducial ``hot'' trans-Alfv\'enic solution presented in Paper
I also exhibits a power-law behavior. Its power-law exponents
can be obtained as a special case ($\beta = 1$) of the preceding results,
corresponding to a parabolic fieldline shape  ($z\propto
\varpi^2$) and $\gamma \propto \varpi$ \citep[see][]{VK01}. As demonstrated
above, our fiducial super-Alfv\'enic flow accelerates significantly more
slowly with distance from the source.

Equation (\ref{scal2}) shows that, as the flow accelerates and
the value of the magnetization function declines, the jet opening half-angle decreases
faster than $\theta$. The cylindrical regime is thus reached at
a finite distance from the source, corresponding to
\begin{equation}\label{efficiency}
\sigma_\infty = \frac{F \sigma_M}{\mu - F \sigma_M}\,,
\quad \gamma_\infty=\frac{\mu}{1 + \sigma_\infty} = \mu - F \sigma_M \,,
\end{equation}
which yield an acceleration efficiency $E_k / {\cal E}_i =
\gamma_\infty / \mu = (1 + \sigma_\infty)^{-1}
= 1- F \sigma_M/\mu $. 
Higher efficiencies thus correspond to larger values of $\mu /F
\sigma_M$. We can use equation (\ref{V_s}) to relate this
parameter combination to the boundary conditions at the base of
the flow: $\mu /F \sigma_M = c \sin \theta_i / V_{p \, i} \sin (\theta_i
-\vartheta_i)$, or, for $\theta_i \approx \pi/2$,
$\mu /F \sigma_M\approx c/ V_{z \, i}$.
The poloidal velocity cannot be very small ($V_p/c \approx
-E/B_\phi$ should be close to 1 to allow the flow to reach large distances; 
see \S~\ref{description}), so a high efficiency requires a large
initial opening half-angle $\vartheta_i$. However, a large value of $\mu /F \sigma_M$
also implies a small value of the exponent $\beta = d \ln
\gamma / d \ln \varpi$, and since $\gamma_\infty$ is always close to $\mu$,
$\varpi_\infty \approx \varpi_i \mu^{1/\beta}$ would have to be large.
Since $z \propto \varpi^{1+\beta}$, this places a practical
upper limit on $\vartheta_i$: it cannot be too large or else the
acceleration would be completed on scales much larger than
typically inferred for GRB outflows ($\sim 10^{14}-10^{15}\, {\rm cm}$).

\citet{DS02} examined a pure-$B_\phi$, strictly radial flow 
without considering the transfield force-balance equation.
They showed that magnetic energy dissipation (modeled in a
parameterized manner) results in efficient acceleration $\gamma \propto r^{0.25}$ and
a final Poynting-to-kinetic energy conversion efficiency $\approx 54 \%$.
Our fiducial solution exhibits a faster acceleration $\gamma \propto r^{0.31}$
and a higher efficiency $\approx 67 \%$.
These results indicate that both dissipation and
the fieldline shape may play an important role in the flow acceleration.

\section{Conclusion}\label{conclusion}
Using the radially self-similar relativistic MHD model presented
in Paper I, we constructed representative ``hot'' and ``cold''
fast-rotator solutions of super-Alfv\'enic flows. We argued that
super-Alfv\'enic outflows in which $|B_\phi/B_p|\gg 1$ already
at the source could plausibly arise in GRB source models
that invoke differential rotation in a disk or a star to account
for the large magnetic field amplitudes that are required for
consistency with the observations. We demonstrated that our
``frozen pulse'' formulation, despite being quasi steady, can still capture the
expected time-variability of the azimuthal field component at
the source. We showed that for typical source parameters the
flows convert Poynting flux to kinetic-energy flux with high efficiency ($
\approx 67\%$ for our
fiducial solution) and collimate to cylindrical structures on
scales $\gtrsim 10^{14}\, {\rm cm}$. Our solutions confirm (some
previous statements in the literature notwithstanding) that
significant magnetic acceleration and collimation can take place at large
distances from the source (well beyond the light cylinder),
although most of the collimation is achieved before the flow
becomes extremely relativistic.

The super-Alfv\'enic solutions derived in this paper are
distinguished from the trans-Alfv\'enic ones obtained in paper I
in two main respects: (1) During the initial thermal-acceleration
phase of the ``hot'' solution, some of the internal energy is transformed into
electromagnetic energy even as another part is used to increase
the poloidal velocity. (The same behavior is exhibited by the
``cold'' solution, except that in that case the energy reservoir
is the bulk initial rotation rather than the thermal energy.)
(2) During the subsequent magnetic-acceleration phase, the rate
of increase of the Lorentz factor with distance from the source
can be significantly lower than in the trans-Alfv\'enic case; the rate of
increase of the jet radius with distance is correspondingly higher. 
Overall, however,, the two types of solution are quite
similar, and we derived analytic scaling relations that describe
them both (see also \citealt{VK01}). We conclude that source
configurations with either $|B_p/B_\phi| \gtrsim 1$ or
$|B_p/B_\phi|\ll 1$ could in principle produce viable GRB
outflows. Another potentially important aspect of such outflows,
namely, the possibility that their initial composition is highly
neutron-rich, can also be  modeled within the theoretical framework that we have
constructed. This is discussed in a separate publication \citep{VPK03}.
\acknowledgements
This work was supported in part by NASA grant
NAG5-12635 and by the U.S. Department of Energy under grant B341495
to the Center for Astrophysical Thermonuclear Flashes at the
University of Chicago. N. V. also acknowledges support from a McCormick Fellowship at
the Enrico Fermi Institute.

\appendix 
\section{Physical Quantities}\label{appA}
In the super-Alfv\'enic regime ($M^2 \gg 1-x^2 $, $x\gg x_A$) of the $r$ self-similar model 
described in Paper I, the expressions for the various physical quantities as functions
of $r$, $\theta$, and $s \equiv ct-\ell$ take the form:
\begin{mathletters}\label{eqs_s}
\begin{eqnarray}
{\mathbf{B}}=
\frac{B_0(s_0)\varpi_0^{2-F}(s_0) x_A^F(s_0)}{g^{F/2}(s_0)}
\left[
\frac{(r \sin \theta)^{F-1}}{{\cal X}^F r \sin(\theta-\vartheta)}
(\hat{z} \cos \vartheta + \hat{\varpi} \sin \vartheta ) -
%\frac{{\cal X}^2 g(s)-x_A^2(s)}{{\cal X}^2 g(s)} 
%\frac{({\cal X}^2 +{\cal M}^2) g(s)}{({\cal X}^2 +{\cal M}^2) g(s)-1}
\frac{\mu(s_0)}{\sigma_M(s_0)} \frac{(r \sin \theta)^{F-2} g^{1/2}}{F {\cal X}^{F-3}
({\cal X}^2 +{\cal M}^2)} \hat{\phi}
\right]\ ,
\label{B_s}
\end{eqnarray}
\begin{eqnarray}
{\mathbf{E}}=
\frac{B_0(s_0)\varpi_0^{2-F}(s_0) x_A^F(s_0)}{g^{F/2}(s_0)}  
\frac{(r \sin \theta)^{F-2} \sin \theta g^{1/2}}{{\cal X}^{F-1} \sin(\theta-\vartheta)}
(\hat{z} \sin \vartheta - \hat{\varpi} \cos \vartheta ) \, ,
\label{E_s}
\end{eqnarray}
\begin{eqnarray}
\frac{{\mathbf{V}}}{c}=
\frac{F \sigma_M(s_0)}{\mu(s_0)} \frac{\sin \theta}{\sin(\theta-\vartheta)}
\frac{{\cal X}^2 +{\cal M}^2}{{\cal X}^2}
(\hat{z} \cos \vartheta + \hat{\varpi} \sin \vartheta ) 
+\frac{x_A^2(s)}{{\cal X} g^{1/2}} 
\hat{\phi}\, ,
\label{V_s}
\end{eqnarray}
\begin{eqnarray}
\xi \gamma = \mu(s) \frac{{\cal M}^2}{{\cal X}^2 +{\cal M}^2}\ ,
\label{xigamma_s}
\end{eqnarray}
\begin{eqnarray}
\gamma \rho_0 =
\left( \frac{B_0(s_0)\varpi_0^{2-F}(s_0) x_A^F(s_0)}{g^{F/2}(s_0)} \right)^2
\frac{\mu(s_0)}{\sigma_M(s_0)} \frac{g(s)}{\sigma_M(s)}
\frac{(r \sin \theta)^{2(F-2)}}{4 \pi c^2 F^2 {\cal
X}^{2(F-2)}({\cal X}^2 +{\cal M}^2)}\ ,
\label{rho_s}
\end{eqnarray}
where $x^2={\cal X}^2 g(s)$, $M^2={\cal M}^2 g(s)$,
$\frac{B_0(s)\varpi_0^{2-F}(s) x_A^F(s)}{g^{F/2}(s)} =
\frac{B_0(s_0)\varpi_0^{2-F}(s_0) x_A^F(s_0)}{g^{F/2}(s_0)}$,
$\mu(s)=\sigma_M(s) \frac{\mu(s_0)}{\sigma_M(s_0)}$,
and the functions $g(s)$, $x_A(s)$, $\sigma_M(s)$, and $q(s)$
can be specified freely.
The functions ${\cal X}^2 (\theta)$, ${\cal M}^2 (\theta)$, and $\vartheta(\theta)$ 
can be found from the integration of the ordinary differential
equations (I.B2d) and (I.B2e) of Paper I,
whereas $\xi$ is given from
\begin{eqnarray}\label{xi_s}
\frac{\xi}{\left(\xi-1\right)^{\frac{1}{\Gamma-1}}} = \frac{g(s)}{q(s)} {\cal M}^2 \,.
\end{eqnarray}
\end{mathletters}


\begin{thebibliography}{} 
\bibitem[Abbett \& Fisher(2003)]{AF03}
Abbett, W. P., \& Fisher, G. H. 2003, \apj, 582, 475

\bibitem[Akiyama et al.(2003)]{A03}
Akiyama, S., Wheeler, J. C., Meier, D., \& Lichtenstadt,
I. 2003, \apj, 584, 954

\bibitem[Balbus \& Hawley(1998)]{BH98}
Balbus, S. A., \& Hawley, J. F. 1998, Rev. Mod. Phys., 70, 1

\bibitem[Bogovalov(2001)]{B01} Bogovalov, S. V. 2001, \aap, 371, 1155

\bibitem[Brummell, Cline, \& Cataneo(2002)]{BCC02}
Brummell, N., Cline, K., \& Cattaneo, F. 2002, \mnras, 329, L73

\bibitem[Cao \& Spruit(1994)]{CS94}
Cao, X., \& Spruit, H. C. 1994, \aap, 287, 80

\bibitem[Chakrabarti \& D'Silva(1994)]{CD94}
Chakrabarti, S. K., \& D'Silva, S. 1994, \apj, 424, 138

\bibitem[Chiueh et al.(1991)Chiueh, Li, \& Begelman(1991)]{CLB91}
Chiueh, T., Li, Z.-Y., \& Begelman, M.C. 1991, \apj, 377, 462

\bibitem[Chiueh et al.(1998)Chiueh, Li, \& Begelman]{CLB98}
Chiueh, T., Li, Z.-Y., \& Begelman, M.C. 1998, \apj, 505, 835

\bibitem[Contopoulos(1995)]{C95} Contopoulos, J. 1995, \apj, 450, 616

\bibitem[Coroniti(1981)]{C81}
Coroniti, F. V. 1981, \apj, 244, 587

\bibitem[Drenkhahn \& Spruit(2002)]{DS02}
Drenkhahn, G., \& Spruit, H. C. 2002, \aap, 391, 1141

\bibitem[Eardley \& Lightman(1975)]{EL75}
Eardley, D. M., \& Lightman, A. P. 1975, \apj, 200, 187

\bibitem[Fryer(1999)]{F99}
Fryer, C. L. 1999, \apj, 522, 413

\bibitem[Haswell, Tajima, \& Sakai(1992)]{HTS92}
Haswell, C. A., Tajima, T., \& Sakai, J.-I. 1992, \apj, 401, 495

\bibitem[Klu\'zniak \& Ruderman(1998)]{KR98}
Klu\'zniak, W., \& Ruderman, M. 1998, \apj, 505, L113

\bibitem[Kobayashi, Piran, \& Sari(1997)]{KPS97}
Kobayashi, S., Piran, T., \& Sari, R. 1997, \apj, 490, 92

\bibitem[Lee, Wijers, \& Brown(2000)]{LWB00}
Lee, H. K., Wijers, R. A. M. J., \& Brown, G. E. 2000, Phys. Rep., 325,
83

\bibitem[Lyutikov \& Blandford(2002)]{LB02}
Lyutikov, M., \& Blandford, R. 2002, preprint (astro-ph/0210671)

\bibitem[MacFadyen \& Woosley(1999)]{MW99}
MacFadyen, A. I., \& Woosley, S. E. 1999, \apj, 524, 262

\bibitem[MacFadyen, Woosley, \& Heger(2001)]{MWH01}
MacFadyen, A. I., Woosley, S. E., \& Heger, A. 2001, \apj, 550, 410

\bibitem[M\'{e}sz\'{a}ros \& Rees(1997)]{MR97}
M\'{e}sz\'{a}ros, P., \& Rees, M. J. 1997, \apj, 482, L29

\bibitem[Miller \& Stone(2000)]{MS00}
Miller, K. A., \& Stone, J. M. 2000, \apj, 534, 398

\bibitem[Okamoto(2002)]{O02} Okamoto, I. 2002, \apjl, 573, L31

\bibitem[Ruderman et al.(2000)Ruderman, Tao, \& Klu\'zniak]{RTK00}
Ruderman, M. A., Tao, L., \& Klu\'zniak, W. 2001, \apj, 542, 243

\bibitem[Shibata, Tajima, \& Matsumoto(1990)]{STM90}
Shibata, K., Tajima, T., \& Matsumoto, R. 1990, \apj, 350, 295

\bibitem[Stella \& Rosner(1984)]{SR84}
Stella, L., \& Rosner, R. 1984, \apj, 277, 312

\bibitem[Thompson \& Murray(2001)]{TM01}
Thompson, C., \& Murray, N. 2001, \apj, 560, 339

\bibitem[Vietri \& Stella(1998)]{VS98}
Vietri, M., \& Stella, L. 1998, \apj, 507, L45

\bibitem[Vlahakis(2003)]{V03}
Vlahakis, N. 2003, \apj, submitted

\bibitem[Vlahakis at al.(2003)Vlahakis, Peng, \& K\"onigl]{VPK03}
Vlahakis, N., Peng, F., \& K\"onigl, A. 2003, \apj, to be submitted

\bibitem[Vlahakis \& K\"onigl(2001)]{VK01}
Vlahakis, N., \& K\"onigl, A. 2001, \apjl, 563, L129

\bibitem[Vlahakis \& K\"onigl (2003)]{VK03}
Vlahakis, N., \& K\"onigl, A. 2003, \apj, submitted, preprint astro-ph/0303482 (Paper I)

\end{thebibliography}
\end{document}